\documentclass[conference]{IEEEtran}
\IEEEoverridecommandlockouts

\usepackage{cite}
\usepackage{amsmath,amssymb,amsfonts}
\usepackage{algorithmic}
\usepackage{graphicx}
\usepackage{textcomp}
\usepackage{xcolor}
\usepackage{booktabs}
\usepackage{array}
\usepackage{tabularx}
\usepackage{multirow}
\usepackage{hyperref}
\usepackage{url}

\usepackage{tikz}
\usetikzlibrary{shapes.geometric, arrows.meta, positioning, fit, backgrounds}
\usepackage[capitalize,noabbrev]{cleveref}

\usepackage[most]{tcolorbox}
\usepackage{balance}
\definecolor{takeawayBorder}{HTML}{1F4E79}
\definecolor{takeawayBack}{HTML}{EAF1F8}
\definecolor{stageId}{HTML}{4A7AB7}
\definecolor{stageScreen}{HTML}{7FA6D9}
\definecolor{stageElig}{HTML}{A8C5E5}
\definecolor{stageIncl}{HTML}{C5DBED}

\newtcolorbox{takeaway}{
    enhanced, unbreakable,
    colback=takeawayBack, colframe=takeawayBorder,
    boxrule=0.5pt, arc=2pt,
    left=5pt, right=5pt, top=2pt, bottom=2pt,
    fontupper=\footnotesize
}

\hypersetup{
    colorlinks=true,
    linkcolor=black,
    citecolor=black,
    urlcolor=blue,
    pdftitle={Measuring What Matters in the Age of AI: The Research Software Metrics Landscape},
    pdfauthor={Addi Malviya-Thakur and Reed Milewicz and Gregory R. Watson and Audris Mockus},
    pdfkeywords={research software, scientific software, software metrics, metrics landscape, FAIR software, software citation, responsible metrics, AI-assisted development, research software engineering, systematic mapping protocol}
}

\def\BibTeX{{\rm B\kern-.05em{\sc i\kern-.025em b}\kern-.08em
    T\kern-.1667em\lower.7ex\hbox{E}\kern-.125emX}}

\usepackage{enumitem}
\setlist{nosep,topsep=2pt,partopsep=0pt}

\begin{document}

\title{Measuring What Matters in the Age of AI:\\
The Research Software Metrics Landscape}

\author{%
\IEEEauthorblockN{%
Addi Malviya-Thakur\textsuperscript{1,2},
Reed Milewicz\textsuperscript{3},
Gregory R. Watson\textsuperscript{1},
Audris Mockus\textsuperscript{2}%
}

\IEEEauthorblockA{%
 \textsuperscript{1}Oak Ridge National Laboratory, USA \quad
 \textsuperscript{2}University of Tennessee, Knoxville, USA \quad
 \textsuperscript{3}Sandia National Laboratories, USA \\
\{malviyaa, watsongr\}@ornl.gov, rmilewi@sandia.gov, audris@utk.edu%
 }%
}

\maketitle

\begin{abstract}
Research software now serves as a primary instrument of science, but the methods the community uses to judge its quality, sustainability, reuse, and impact remain scattered across software engineering, open science, FAIR software, citation, and research-assessment literature. AI-assisted development sharpens the problem: code can be produced at rates that outpace human review capacity, provenance and attribution are harder to track, scholarly credit systems cannot keep pace, and existing metrics inherit assumptions about how research software gets developed that may no longer apply. To aid in navigating these challenges, we present a focused landscape synthesis of research software metrics based on 49 substantive references, organizing the recurring metrics into eight families described by data source, claimed construct, evidence of validity, automation potential, and susceptibility to misuse. Many widely cited research software metrics serve as partial signals but make weak standalone measures of scientific impact. The AI era opens or widens gaps around provenance, maintainability of generated code, reviewer effort, model and data dependencies, reuse and community-health signal inflation, and credit for human labor that AI does not perform. We argue for responsible, contextual, mixed-method use of research software metrics, and offer decision guidance for Research Software Engineers (RSEs), principal investigators, institutions, and funders selecting metrics that support better software, fairer credit, and more trustworthy science. All findings rest on the 49-reference focused review; a reproducible search-and-triage protocol and a 727-record triaged corpus accompany the paper as a transparency artifact for a future full systematic mapping study.
\end{abstract}


\vspace{-6pt}
\section{Introduction}
\vspace{-4pt}
\label{sec:introduction}
Research software now serves as a primary instrument of science. Simulations, data pipelines, machine learning models, and analysis scripts mediate hypotheses, evidence, and conclusions across nearly every research domain~\cite{howison2016software,hettrick2014uk,goble2014better}. While reliance on this software has grown, the metrics available to evaluate it across multiple dimensions have not kept pace. Research Software Engineers (RSEs) work to improve reproducibility, reliability, usability, and sustainability, but these contributions are often represented in evaluation through weak proxies such as repository stars, download counts, publication citations, or anecdotal user testimonials~\cite{cosden2023rse,katz2019community}.

AI-assisted development sharpens the problem. Copilot- and large language model (LLM)-based tooling raises the rate at which research code can be generated, modified, and documented relative to the capacity of human review, maintenance, and scholarly credit systems to absorb it~\cite{chen2021codex,perry2023copilot,radar26,aiCodeReview26}. Industrial reports document the transfer of effort from writing to reviewing at scale: at Meta, per-developer diff volume rose by half and per-diff code volume more than doubled year-over-year, with agentic AI responsible for the majority of that growth, forcing organizations to build risk-triage systems that auto-approve low-risk changes so that human reviewers can concentrate on the rest~\cite{radar26,RevRisk26}. AI-assisted coding changes the economics of producing research software, but not the need for evidence about whether software is reliable, reusable, maintainable, citable, secure, or scientifically trustworthy. We treat metrics in this setting not as administrative indicators but as \emph{evidence infrastructure} for assessing essential software characteristics.

Existing work on research software metrics is fragmented. The Findable, Accessible, Interoperable, Reusable (FAIR) Principles for Research Software (FAIR4RS)~\cite{barker2022fair4rs,lamprecht2020fair} adapt these properties to software. Scholarly credit has been explored in the FORCE11 Software Citation Principles~\cite{smith2016force11}, while sustainability and community-health guidance comes from the Software Sustainability Institute and the Research Software Alliance~\cite{resa,ssi}. The Declaration on Research Assessment (DORA)~\cite{dora} and the Leiden Manifesto~\cite{hicks2015leiden} push back against narrow, citation-only assessment. 
Software engineering offers a long tradition of code-quality metrics~\cite{fenton2014software,kitchenham2007guidelines}, much of it developed outside research contexts. RSEs, principal investigators (PIs), institutions, and funders selecting metrics today face the burden of synthesizing across these broad topics without a shared landscape; this paper
 aims to fill 
 that gap.

Drawing on a focused review of 49 substantive references across software engineering, open science, FAIR software, citation, and research-assessment literature, we organize the recurring metrics into eight families, indexed by data source, claimed construct, evidence of validity, automation potential, and risk of misuse. We frame the work deliberately as a landscape rather than a ranking. Our goal is not to create a single metric for measuring the impact of research software, but to help practitioners choose metrics that fit their purpose, context, and stakeholders, and that ultimately can be used to improve the software and the way that it is developed. We pay attention to gaps the AI era opens or sharpens: code provenance, maintainability of generated artifacts, reviewer effort, and credit for human labor that AI systems do not perform.
We organize our synthesis around five research questions (RQs):

\begin{enumerate}[leftmargin=1cm]
    \item[\textbf{RQ1.}] What \emph{metric families} do researchers use or propose to evaluate research software?
    \item[\textbf{RQ2.}] What data sources feed the metrics in these families?
    \item[\textbf{RQ3.}] What constructs do these metrics claim to measure, and how do indicators relate to constructs?
    \item[\textbf{RQ4.}] Which metrics are validated, which serve as proxies, and which are vulnerable to gaming or bias?
    \item[\textbf{RQ5.}] What new metric gaps appear in the age of AI-assisted research software?
\end{enumerate}

The novelty of this work is not the existence of individual metric families, which are well established, but a \emph{decision-oriented landscape} that separates indicators from the constructs they claim to measure, compares validation strength and misuse risk across families, and identifies where AI-assisted development strains the assumptions behind common metrics. Concretely, this paper makes four contributions: (1) a structured landscape of research software metrics organized into eight recurring families and indexed against software quality, sustainability, FAIR principles, reuse, citation, community health, security, and environmental concerns (\Cref{sec:taxonomy}); (2) an assessment of which metrics carry validation evidence, which serve as proxies, and which are vulnerable to gaming or bias (\Cref{sec:evidence}); (3) an AI-era gap analysis identifying where existing measures inherit assumptions that AI-assisted development strains (\Cref{sec:discussion}); and (4) decision guidance for RSEs, principal investigators, institutions, and funders selecting metrics in context, including a compact use-case decision guide (\Cref{sec:implications}, \Cref{tab:decision-guide}). Supporting these contributions, we additionally release a reproducible search-and-triage protocol with a 727-record triaged corpus as a transparency artifact for follow-on work; the claims of this paper rest on the focused review alone (\Cref{sec:method}).

\vspace{-4pt}
\section{Background and Definitions}
\vspace{-4pt}
\label{sec:background}

We begin by defining research software, distinguishing constructs from metrics, summarizing the responsible-metrics tradition, reviewing FAIR software and citation standards, and introducing AI-assisted research software development.

\vspace{-4pt}
\subsection{Research Software}
\vspace{-4pt}
We adopt a broad working definition aligned with FAIR4RS~\cite{barker2022fair4rs} and the Research Software Alliance~\cite{resa}: \emph{research software} comprises source code, algorithms, scripts, computational workflows, executables, notebooks, and configuration
created during the research process or for a research purpose.
The definition spans general-purpose scientific libraries, domain-specific simulation packages, individual analysis scripts, container recipes, and AI training and inference pipelines. We chose an inclusive definition deliberately: research software ranges from one-author exploratory scripts to multi-institution platforms with thousands of contributors, and a single metric rarely fits all cases~\cite{hannay2009scientists,hettrick2014uk,carver2022survey}.

\vspace{-4pt}
\subsection{Metrics, Indicators, and Constructs}
\vspace{-4pt}
Following standard measurement terminology~\cite{fenton2014software,hicks2015leiden}, we distinguish three concepts:
\begin{itemize}
    \item A \textbf{construct} is an abstract property of interest, such as maintainability, scientific impact, or sustainability.
    \item An \textbf{indicator} is an observable signal believed to relate to a construct (e.g. a commit cadence as an indicator of maintenance activity).
    \item A \textbf{metric} is a defined, repeatable computation over data sources that yields an indicator value.
\end{itemize}
This distinction matters because the three are easily conflated in practice: ``GitHub stars'' is a metric, but whether stars validly \emph{indicate} scientific impact or reuse is a question the metric itself cannot answer (\Cref{sec:evidence}).
\vspace{-4pt}
\subsection{Responsible Metrics}
\vspace{-4pt}
DORA~\cite{dora}, the Leiden Manifesto~\cite{hicks2015leiden}, and the Metric Tide report~\cite{wilsdon2015metric} converge on principles for \emph{responsible} metric use. These frameworks hold that quantitative evaluation should support rather than supplant qualitative judgment, that the data and methods behind a metric must be transparent and appropriate to context, and that those evaluated should be able to verify the underlying data. We apply this framing throughout: a metric is not ``good'' or ``bad'', 
only more or less appropriate for a given purpose, stakeholder, and context.

\vspace{-4pt}
\subsection{FAIR Software and Software Citation}
\vspace{-4pt}
Two community standards inform much of the current work in the area. The FAIR4RS Principles~\cite{barker2022fair4rs,lamprecht2020fair} adapt FAIR data principles to software and emphasize persistent identifiers, rich metadata, open licensing, and versioning. The FORCE11 Software Citation Principles~\cite{smith2016force11} establish software as a legitimate scholarly product worthy of citation, with associated infrastructure including the Citation File Format (CFF)~\cite{druskat2021cff}, Zenodo--GitHub integration, and the Journal of Open Source Software (JOSS)~\cite{smith2018joss}. The community evaluates these standards using metrics themselves (e.g. the share of repositories with a CITATION.cff file), so FAIR and citation infrastructure serve as both object and source of measurement.

\vspace{-4pt}
\subsection{AI-Assisted Research Software Development}
\vspace{-4pt}
Multiple empirical studies document the capabilities, usability, and risks of AI coding assistants in software development broadly~\cite{chen2021codex,perry2023copilot,vaithilingam2022expectation}. The change is not only quantitative
but also qualitative. As we argue in \Cref{sec:discussion}, provenance becomes harder to trace when commits do not record which model or prompt produced the code~\cite{chen2021codex,druskat2021cff}, code review becomes the scarce resource as production accelerates~\cite{perry2023copilot,vaithilingam2022expectation,radar26,aiCodeReview26}, and the boundary between human and machine authorship blurs in ways that existing software-citation infrastructure does not record~\cite{smith2016force11,druskat2021cff}. Industrial-scale evidence has begun to accumulate: automated code-review and reviewer-recommendation systems are now being deployed to absorb AI-driven diff volume that human review cannot, transferring reviewer effort onto ML risk models and LLM-based checks rather than eliminating it~\cite{radar26,RevRisk26}. Metric systems designed for human-authored software encode assumptions about commits as work units, contributor identity, and test authorship that AI-assisted development strains.

\vspace{-4pt}
\subsection{Prior Reviews and Surveys}
\vspace{-4pt}
Empirical practice surveys anchor the field: Hannay et al.~\cite{hannay2009scientists} surveyed 1{,}972 scientists in 2009, and Carver et al.~\cite{carver2022survey} updated that picture for the US-RSE community in 2022. Adjacent literature reviews include Heaton and Carver~\cite{heaton2015claims} on software engineering (SE) practices in science and Arvanitou et al.~\cite{arvanitou2021mapping} on SE practices in scientific software. The closest direct precursor is Eisty et al. ~\cite{eisty2018metrics}, who surveyed metric use among research software developers. Our approach is instead to compare metrics across families and ask how AI changes what they measure, rather than surveying use in current practice.
Community standards reviews including Lamprecht et al.~\cite{lamprecht2020fair} (FAIR) and Howison and Bullard~\cite{howison2016software} (software mention) cover specific facets.

Recent US-RSE proceedings have used research software metrics in specific contexts: complexity-metric-aware code reviews for computational science and engineering (CSE) software~\cite{willenbring2023complexity}, RSE impact assessed through coauthorship and collaboration networks~\cite{reidy2023impact}, and empirical surveys of research software on GitHub at US research universities, first at R1 institutions~\cite{schwartz2024github} and later extended to R2 and doctoral/professional universities~\cite{schwartz2025r2}. The present paper differs by synthesizing metric families across software quality, FAIRness, sustainability, citation, community health, security, and emerging AI-era concerns within a single practitioner-facing landscape, rather than instrumenting a single context, and by asking how AI-assisted development changes the interpretation of the metrics those papers and others rely on.

\vspace{-4pt}
\section{Method}
\label{sec:method}
\vspace{-4pt}
We drew on established guidance for systematic reviews and mapping studies in software engineering~\cite{kitchenham2007guidelines,petersen2015guidelines,petersen2008systematic} and adapted reporting conventions informed by the Preferred Reporting Items for Systematic Reviews and Meta-Analyses (PRISMA)~\cite{page2021prisma}. This paper is a structured landscape synthesis based on a focused review of 49 substantive references. Separately, we contribute a reproducible search-and-triage protocol that has produced a 727-record triaged corpus, prepared for a future full systematic mapping study. A landscape synthesis suits the goal of identifying coverage, gaps, and clusters in a heterogeneous body of work that does not lend itself to effect-size synthesis~\cite{petersen2008systematic}. We adopted deductive-plus-inductive coding and explicit inclusion and exclusion criteria as recommended by the SE mapping-study guidance above, with recent adjacent reviews~\cite{heaton2015claims,arvanitou2021mapping} as exemplars.
We do not claim this focused review is an exhaustive systematic mapping study; its evidentiary target is saturation at the metric-family level rather than prevalence estimation. We use the 49-source set to identify recurring families and construct-validity issues, reserving frequency, coverage, and inter-rater agreement claims for the future 727-record mapping study.

\vspace{-4pt}
\subsection{Focused Review}
\vspace{-4pt}
We define a \emph{substantive reference} as one that introduces, defines, validates, critiques, or aggregates research-software metrics directly, as opposed to citing them only in passing. We excluded sources that addressed software metrics without a research-software context and sources that mentioned metrics incidentally to a different argument. We seeded the set from our working knowledge of research-software metrics and adjacent literature, then expanded it by forward and backward snowballing from eight foundational papers that anchor the relevant prior work~\cite{eisty2018metrics,heaton2015claims,arvanitou2021mapping,barker2022fair4rs,hannay2009scientists,howison2016software,kabir2024llm,wilson2017good}. We acknowledge the resulting seed bias and plan to address it through the pilot-validation and member-check mechanisms in a future extension.

\begin{figure}[!t]
\centering
\begin{tikzpicture}[scale=0.88, transform shape,
    node distance=2.5mm,
    every node/.style={font=\footnotesize},
    bigbox/.style={
        rectangle, draw=takeawayBorder, rounded corners=3pt, line width=0.6pt,
        text width=0.74\columnwidth, minimum height=7mm,
        align=center, inner sep=3pt
    },
    smbox/.style={
        rectangle, draw=takeawayBorder, rounded corners=3pt, line width=0.5pt,
        text width=0.74\columnwidth, minimum height=6mm,
        align=center, inner sep=2pt, font=\scriptsize
    },
    sidebox/.style={
        rectangle, draw=takeawayBorder!70, rounded corners=2pt, line width=0.4pt,
        text width=0.30\columnwidth, minimum height=5mm,
        align=center, inner sep=2pt, font=\scriptsize\itshape, fill=stageIncl!22
    },
    arr/.style={-{Latex[length=1.6mm]}, thick, draw=takeawayBorder},
    arrside/.style={-{Latex[length=1.2mm]}, thick, draw=takeawayBorder!70}
]

\node[smbox, fill=stageId!35] (seeds)
    {\textbf{Seed set} from working knowledge of research-software metrics};

\node[bigbox, below=of seeds, fill=stageId!50] (anchors)
    {\textbf{Eight foundational anchors}\\
     {\scriptsize Eisty, Heaton, Arvanitou, Barker (FAIR4RS), Hannay,\\
      Howison, Kabir, Wilson}};

\node[bigbox, below=of anchors, fill=stageScreen!35] (snow)
    {\textbf{Forward + backward snowballing}\\
     across software engineering, open science, FAIR software,\\
     citation, and research-assessment literatures};

\node[sidebox, right=4mm of snow] (cross)
    {Cross-checks:\\
     FAIR4RS\\
     CHAOSS\\
     RSQKit};

\node[bigbox, below=of snow, fill=stageIncl!60] (corpus)
    {\textbf{49 substantive references}\\
     {\scriptsize that introduce, define, validate, critique,\\
      or aggregate research-software metrics directly}};

\node[bigbox, below=of corpus, fill=stageElig!45] (outputs)
    {Eight recurring metric families (\Cref{sec:taxonomy})\\
     Evidence and validation analysis (\Cref{sec:evidence})\\
     AI-era gap analysis (\Cref{sec:discussion})};

\draw[arr] (seeds) -- (anchors);
\draw[arr] (anchors) -- (snow);
\draw[arr] (snow) -- (corpus);
\draw[arr] (corpus) -- (outputs);
\draw[arrside] (cross.west) -- (snow.east);

\end{tikzpicture}
\vspace{-12pt}
\caption{Focused review flow: from seed set to 49 substantive references, cross-checked against community resources.}
\label{fig:focused-review}
\end{figure}
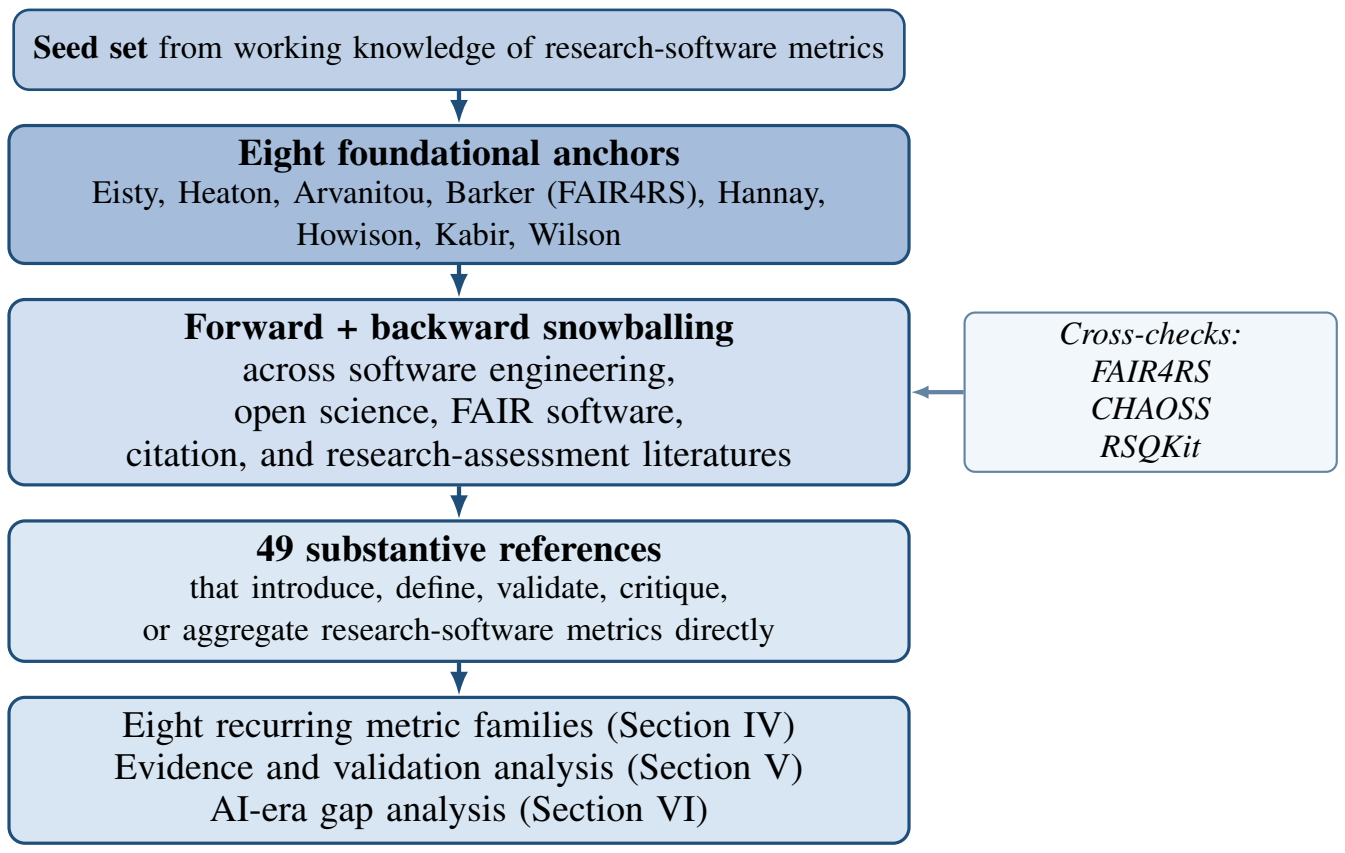

Additional candidate sources were surfaced during seeding and snowballing but excluded when they did not meet the substantive-reference criterion. We cross-checked coverage against FAIR4RS, the Community Health Analytics for Open Source Software (CHAOSS) metrics~\cite{chaoss}, and the Research Software Quality Kit (RSQKit) resources~\cite{rsqkit}; these checks did not surface families outside the eight we identified, though they should be understood as a structured spot-check against three widely cited community resources rather than as an exhaustive survey of every framework in the area. As a concrete example of how the cross-checks worked, CHAOSS's time-to-first-response metric, surfaced during the check, mapped cleanly onto the existing community-health family, and RSQKit's documentation-quality indicators mapped onto software quality; neither required a new family. We treated family-level saturation as reached when additional snowballing and cross-checking produced examples assignable to the existing families rather than candidate new families. \Cref{fig:focused-review} summarizes the review flow. We derived initial codes deductively from the RQs,
refined them inductively from the reviewed sources, and resolved family boundaries and edge cases through author discussion. We do not report inter-rater reliability statistics at the focused-review stage; the future planned mapping study upgrades to two reviewers with Cohen's $\kappa$.

\smallskip
\noindent\emph{Why 49 sources rather than the 727-record corpus.} The two evidence sets arise from separate processes. The 49 references were selected by the seeding-and-snowballing process above under the substantive-reference criterion, stopping at family-level saturation; they were \emph{not} sampled from the 727-record corpus, and no target count was fixed in advance. The 727-record corpus is the output of the separate search-and-triage protocol and the designated input for the planned mapping study. Because snowballing and the cross-checks stopped producing candidate new families, we expect that pass to \emph{refine} the landscape rather than restructure it: prevalence and coverage estimates, formal inter-rater agreement, and possibly splits or merges at family boundaries, with genuinely new families possible but unlikely.

\begin{table}[!t]
\caption{Data-extraction schema for the focused review.}
\vspace{-6pt}
\label{tab:extraction}
\centering
\scriptsize
\renewcommand{\arraystretch}{1.05}
\begin{tabularx}{\columnwidth}{@{}lX@{}}
\toprule
\textbf{Field} & \textbf{Description} \\
\midrule
Citation & Full citation or identifier \\
Year & Publication year \\
Venue type & Journal, conference, preprint, community report \\
RS definition & How the source defines research software \\
Metric family & One of the eight families developed inductively in \Cref{sec:taxonomy} \\
Specific metric & The actual measure or indicator proposed \\
Data source & Git, registry, publication metadata, survey, CI logs \\
Stakeholder & RSE, PI, institution, funder, maintainer, or user \\
Claimed construct & What the metric is supposed to measure \\
Validation & Whether and how the metric was validated \\
Automation & Manual, semi-automated, or automated \\
Risks & Bias, gaming, incompleteness, inequity, or false precision \\
AI relevance & Whether the metric applies to AI-assisted software \\
\bottomrule
\end{tabularx}
\end{table}

\vspace{-4pt}
\subsection{Reproducibility Artifact}
\vspace{-4pt}
The replication package\footnote{\url{https://doi.org/10.5281/zenodo.22809837}}
is organized into two scopes. The focused-review scope holds the 49 substantive references with family assignments, the per-source extraction over the \Cref{tab:extraction} fields, the \Cref{tab:families}, \Cref{tab:matrix}, and \Cref{tab:ai-gaps} content as CSV, the family codebook, and the extraction-schema documentation; remaining bibliography entries support policy, methodology, and community context. 
The search-and-triage scope holds a reproducible OpenAlex~\cite{priem2022openalex} pipeline prepared for a planned journal-length systematic mapping study; the present paper's claims rest on the focused review, not on the search-and-triage corpus.
Six query families paired research-software terms with metric, quality/sustainability, RSE impact, citation/credit, FAIR assessment, and AI-assisted-development/provenance terms; queries plus snowballing produced a 4{,}452-record combined corpus during 2026, automated pre-screening retained 722 records, and five foundational sources were re-added after bibliographic verification, yielding the 727-record triaged corpus. 
The scope also contains the executable notebook, the screening protocol
with its
compensating validity mechanisms,
and a future-work plan.
Follow-on work will run the full two-reviewer screening, family coding, sensitivity analysis, and member-check against the triaged corpus.

\section{The Research Software Metrics Landscape}
\label{sec:taxonomy}
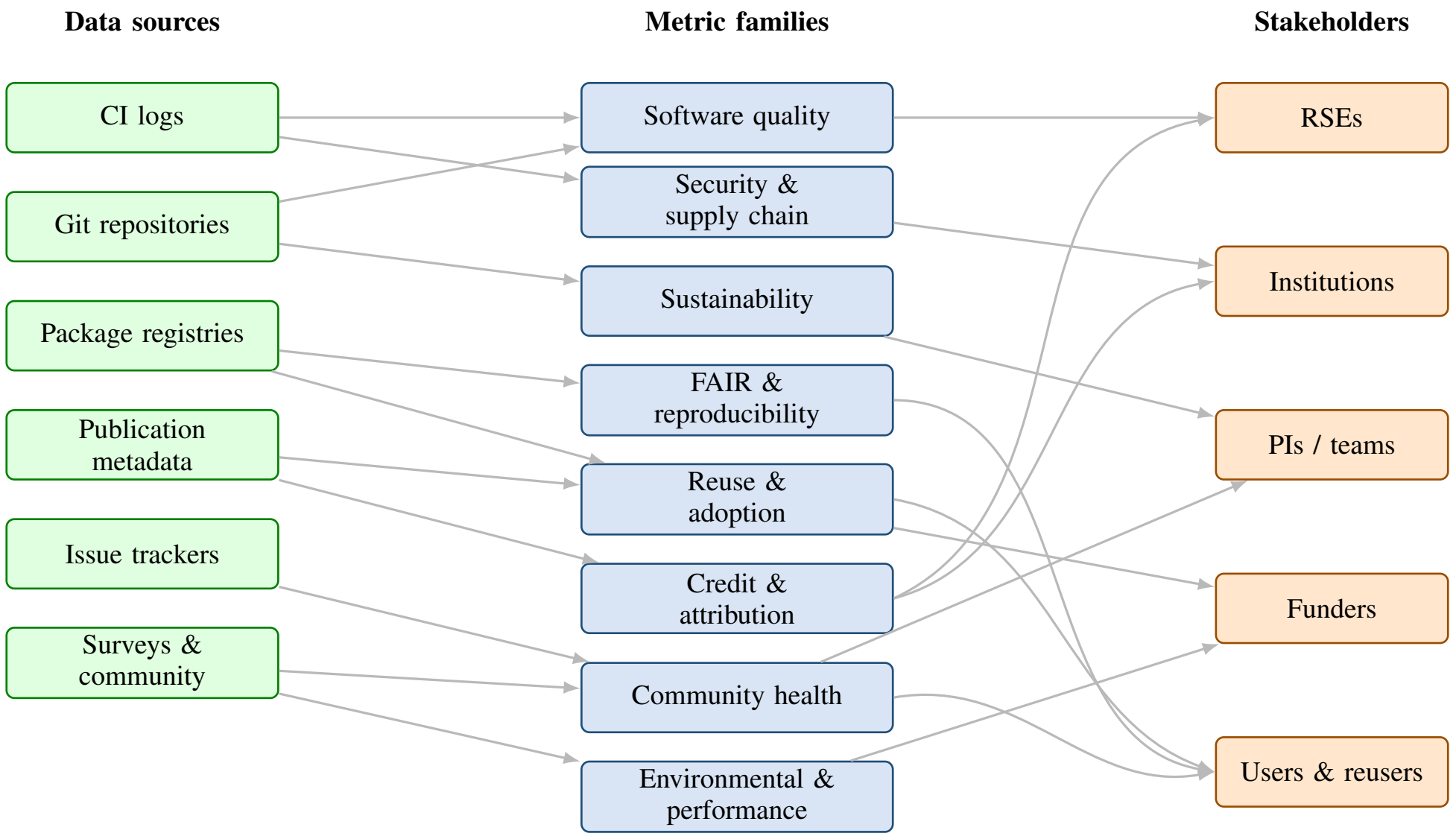
\begin{figure*}[!t]
\centering
\begin{tikzpicture}[scale=0.88, transform shape,
    every node/.style={font=\footnotesize},
    source/.style={
        rectangle, draw=green!50!black, rounded corners=2pt,
        text width=26mm, minimum height=7mm, line width=0.5pt,
        align=center, fill=green!12, inner sep=2pt
    },
    family/.style={
        rectangle, draw=takeawayBorder, rounded corners=2pt,
        text width=30mm, minimum height=7mm, line width=0.5pt,
        align=center, fill=stageScreen!30, inner sep=2pt
    },
    stake/.style={
        rectangle, draw=orange!60!black, rounded corners=2pt,
        text width=22mm, minimum height=7mm, line width=0.5pt,
        align=center, fill=orange!20, inner sep=2pt
    },
    arr/.style={
        -{Latex[length=1.5mm]}, gray!55, semithick
    },
    curvearr/.style={
        -{Latex[length=1.5mm]}, gray!55, semithick
    },
    headerstyle/.style={
        font=\footnotesize\bfseries, anchor=south, align=center
    }
]

\node[headerstyle] at (0, 0.75) {Data sources};
\node[headerstyle] at (6, 0.75) {Metric families};
\node[headerstyle] at (12, 0.75) {Stakeholders};

\node[source] (s1) at (0, 0)    {CI logs};
\node[source] (s2) at (0, -1.1) {Git repositories};
\node[source] (s3) at (0, -2.2) {Package registries};
\node[source] (s4) at (0, -3.3) {Publication\\metadata};
\node[source] (s5) at (0, -4.4) {Issue trackers};
\node[source] (s6) at (0, -5.5) {Surveys \&\\community};

\node[family] (f1) at (6, 0)     {Software quality};
\node[family] (f2) at (6, -0.85) {Security \&\\supply chain};
\node[family] (f3) at (6, -1.85) {Sustainability};
\node[family] (f4) at (6, -2.85) {FAIR \&\\reproducibility};
\node[family] (f5) at (6, -3.85) {Reuse \&\\adoption};
\node[family] (f6) at (6, -4.85) {Credit \&\\attribution};
\node[family] (f7) at (6, -5.85) {Community health};
\node[family] (f8) at (6, -6.85) {Environmental \&\\performance};

\node[stake] (st1) at (12, 0)     {RSEs};
\node[stake] (st2) at (12, -1.65) {Institutions};
\node[stake] (st3) at (12, -3.3)  {PIs / teams};
\node[stake] (st4) at (12, -4.95) {Funders};
\node[stake] (st5) at (12, -6.6)  {Users \& reusers};

\draw[arr] (s1) -- (f1);
\draw[arr] (s1) -- (f2);
\draw[arr] (s2) -- (f1);
\draw[arr] (s2) -- (f3);
\draw[arr] (s3) -- (f4);
\draw[arr] (s3) -- (f5);
\draw[arr] (s4) -- (f5);
\draw[arr] (s4) -- (f6);
\draw[arr] (s5) -- (f7);
\draw[arr] (s6) -- (f7);
\draw[arr] (s6) -- (f8);

\draw[arr] (f1) -- (st1);
\draw[arr] (f2) -- (st2);
\draw[arr] (f3) -- (st3);
\draw[arr] (f5) -- (st4);
\draw[arr] (f7) -- (st3);
\draw[arr] (f8) -- (st4);

\draw[curvearr] (f4.east) to[out=0, in=170] (st5.west);
\draw[curvearr] (f5.east) to[out=-10, in=160] (st5.west);
\draw[curvearr] (f7.east) to[out=10, in=190] (st5.west);

\draw[curvearr] (f6.east) to[out=15, in=-170] (st2.west);
\draw[curvearr] (f6.east) to[out=25, in=-170] (st1.west);

\end{tikzpicture}
\caption{Research software metrics landscape: data sources, metric families, and stakeholders.}
\label{fig:taxonomy}
\end{figure*}

\vspace{-2pt}
The focused review surfaced eight recurring \emph{metric families} that organize the landscape. \Cref{tab:families} introduces each family with its anchor citations, example indicators, claimed construct, validation evidence, and AI-era relevance; \Cref{tab:matrix} complements this with each family's primary data sources, automation potential, and misuse risk. Together the two tables cover the five-axis indexing promised in the abstract.
Throughout the paper, validation evidence (in both the tables and the text) is tagged with one of three tiers: \emph{established} (direct empirical support in the cited literature), \emph{emerging} (partial or adjacent empirical support), or \emph{anticipated} (a projected risk requiring future empirical study). \Cref{fig:taxonomy} visualizes how data sources feed families and which stakeholders the families typically inform, with curved arrows marking cross-row connections; mappings shown are representative of the connections we observed, though in practice the relationships are more many-to-many than any single diagram can convey legibly. Three edges illustrate the cross-family structure. Publication metadata feeds both reuse and credit: the same citation trace is read as adoption evidence by a funder and attribution evidence by an RSE. Package registries feed both FAIR and reuse: the metadata that makes software findable also makes download and dependent-package counts computable, so weaknesses in one family propagate into the other. And credit points to both institutions and RSEs, since attribution infrastructure serves career assessment and individual recognition simultaneously.
The main cross-cutting observation is that the eight families share data sources but differ sharply in validation evidence and in the gap between automation and interpretation; the remainder of this section examines each family.

\begin{table*}[!t]
\caption{Eight metric families: indicators, constructs, validation, and AI relevance.}
\vspace{-6pt}
\label{tab:families}
\centering
\scriptsize
\setlength{\tabcolsep}{4pt}
\renewcommand{\arraystretch}{1.0}
\begin{tabularx}{\textwidth}{@{}l X X X X@{}}

\toprule
\textbf{Family} & \textbf{Example indicators} & \textbf{Claimed construct} & \textbf{Validation evidence} & \textbf{AI-era relevance} \\
\midrule
Software quality &
Test coverage, CI pass rate, code review depth, defect density, modularity, documentation completeness &
Reliability, readability, and maintainability of code &
\textit{Established}~\cite{fenton2014software,wilson2014best,wilson2017good,eisty2018metrics}. Classical SE measures studied for decades; closest survey of RSE metric awareness &
AI-assisted coding can produce code at high volume with uneven reliability and security characteristics~\cite{chen2021codex,perry2023copilot} \\
Sustainability &
Maintainer activity, bus factor~\cite{cosentino2015busfactor}, release cadence, governance maturity, onboarding documentation &
Long-term project survival and maintainer capacity &
\textit{Emerging}~\cite{jansen2014assessing,cosentino2015busfactor,carver2022survey}. Sustainability framing, bus-factor formalization, RSE practice survey evidence &
Creation increasingly AI-assisted; long-term maintenance and stewardship remain predominantly human RSE work~\cite{cosden2023rse} \\
FAIR / reproducibility &
Persistent IDs, metadata richness, license clarity, provenance, versioning, container availability &
Findability, accessibility, interoperability, reusability, and reproducibility &
\textit{Emerging}~\cite{barker2022fair4rs,lamprecht2020fair,larsonneur2018workflows,grayson2023workflows}. Foundational FAIR-for-software principles plus empirical reproducibility evidence &
AI-era workflows 
complicate provenance and reproducibility demands \\
Reuse / adoption &
Downloads, dependent packages, citations, forks, package-manager use &
Scholarly and engineering reuse beyond the producer &
\textit{Established for citations; emerging for software-specific signals}~\cite{howison2016software,du2021software,borges2018githubstars}. Scholarly trace of software reuse plus critique of stars-as-popularity &
Signals whether software travels beyond a single project or paper \\
Credit / attribution &
Software citations, DOIs, CFF presence, contributor roles (CRediT) &
Recognition of code-producing and non-code RSE labor &
\textit{Emerging}~\cite{smith2016force11,druskat2021cff,smith2018joss,credit}. Software citation principles, citation file format, working credit system &
Can make RSE labor more visible in scholarly records where citation practice and contributor-role taxonomies are adopted \\
Community health &
Contributor diversity, response time, governance maturity, mentoring, code-of-conduct presence &
Healthy community capacity and contributor inclusion &
\textit{Emerging}~\cite{chaoss,jansen2014assessing}. Community-derived metric suite plus early adoption-of-metrics framework &
Captures community building and user-support work \\
Security / supply chain &
Dependency freshness, known vulnerabilities, signed releases, SBOM presence, review status &
Supply-chain integrity and attack-surface visibility &
\textit{Established for code; emerging for ML model and data}~\cite{cisa2023sbom,scorecard}. Government and community-led supply-chain assessment work &
AI-generated code may inflate provenance risk by introducing dependencies whose origin and review status are unclear~\cite{perry2023copilot} \\
Environmental / performance &
Runtime, memory, energy use, computational cost per unit of work &
Resource cost and environmental footprint of compute &
\textit{Established for ML workloads; emerging for general software}~\cite{henderson2020greenml,patterson2021carbon}. Foundational energy and carbon measurement work &
AI and large-scale workflows intensify resource and energy questions \\
\bottomrule
\end{tabularx}
\end{table*}

\begin{table*}[tbp]
\caption{Eight metric families: data sources, automation, and misuse risk.}
\vspace{-6pt}
\label{tab:matrix}
\centering
\scriptsize
\setlength{\tabcolsep}{4pt}
\renewcommand{\arraystretch}{1.0}
\newcolumntype{Y}{>{\hsize=.75\hsize}X}%
\newcolumntype{Z}{>{\hsize=1.125\hsize}X}%
\begin{tabularx}{\textwidth}{@{}l Z Y Z@{}}
\toprule
\textbf{Family} & \textbf{Primary data sources} & \textbf{Automation potential} & \textbf{Misuse / bias risk} \\
\midrule
Software quality &
Git history; CI logs; static analyzers; review records &
High; CI and static analysis fully automated &
Structural metrics may pass on AI-generated code that is nonetheless brittle \\
Sustainability &
Contributor history; release timeline; governance docs; funding metadata &
Medium; counts automated, governance assessment manual &
Bus factor of one may be fine for stable scope; survivorship bias \\
FAIR / reproducibility &
Repository metadata; citation files (CFF, DOI); container manifests; registries &
Medium; metadata checks automated, semantic checks manual &
Checkbox compliance without behavior change \\
Reuse / adoption &
Citation databases; dependent-package registries; download stats; mention texts &
High for counts; medium for qualitative mentions &
Stars are not use; downloads are not active reuse; transitive dependencies pollute counts \\
Credit / attribution &
Author lists; CITATION.cff; CRediT taxonomies; JOSS metadata; contributor logs &
Low; taxonomy mapping requires human input &
Authorship inflation; invisible labor remains invisible; AI contribution complicates attribution \\
Community health &
Issue and PR threads; response times; contributor diversity; governance signals &
Medium; counts automated, quality assessment manual &
Bot-driven inflation; activity is not health \\
Security / supply chain &
SBOMs; vulnerability databases; dependency graphs; signed attestations &
High; SBOM generation and dependency scanning automated &
SBOM completeness is not security; check fatigue \\
Environmental / performance &
Pipeline energy meters; compute logs; carbon-aware schedulers &
Medium; energy logging requires instrumentation &
Cherry-picking lightweight runs; gaming via cloud-region selection \\
\bottomrule
\end{tabularx}
\vspace{3pt}
\end{table*}

\vspace{-4pt}
\subsection{Software Quality (RQ1, RQ2)}
\vspace{-4pt}
Quality-oriented metrics include classical software engineering measures such as test coverage, cyclomatic complexity, defect density, code-review participation, and continuous-integration pass rates~\cite{fenton2014software,wilson2014best,wilson2017good}. 
Eisty et al.~\cite{eisty2018metrics} surveyed research software developers specifically on metric use and found that awareness of formal code metrics was uneven and that process metrics were used less than product metrics. In research contexts the community extends classical measures with documentation completeness, example runnability, and tutorial quality~\cite{smith2018joss}. Data sources are predominantly 
version-control host (commit and pull-request metadata), CI logs, and the repository contents themselves. Many quality metrics are well-defined and automate easily, though their predictive validity for downstream maintenance outcomes varies (\Cref{sec:evidence}). They transfer cleanly within a language and domain but poorly across them~\cite{hall2012slr,radjenovic2013metrics}; a one-off exploratory notebook is not poorly engineered for lacking unit tests in the way a community library would be.

\vspace{-6pt}
\subsection{Sustainability (RQ1, RQ2)}
\vspace{-4pt}
Sustainability metrics track whether a project will keep working~\cite{ssi,jansen2014assessing,thakur2025abandoned}. Common indicators include maintainer activity over time, the ``bus factor''~\cite{cosentino2015busfactor}, release cadence, governance documentation, and an onboarding pathway for new contributors; data sources include Git history, governance files, and structured surveys~\cite{carver2022survey}. These measures matter for community libraries and infrastructure, but can misclassify small, scope-stable projects as inactive when low activity reflects feature completeness rather than abandonment.

\vspace{-6pt}
\subsection{FAIR and Reproducibility (RQ1, RQ2)}
\vspace{-4pt}
The FAIR4RS Principles~\cite{barker2022fair4rs,lamprecht2020fair} generate a checklist-style metric family: persistent identifiers, machine-readable metadata, open licenses, semantic versioning, and dependency declarations. Reproducibility-adjacent indicators include reproducible-environment specifications (Dockerfiles, Conda, Nix) and containerized workflow definitions (CWL, Nextflow, Snakemake)~\cite{larsonneur2018workflows,grayson2023workflows}. These metrics signal intent, but as with all checklists they can be satisfied superficially without producing actually reusable software.

\vspace{-4pt}
\subsection{Reuse and Adoption (RQ1, RQ2, RQ3)}
\vspace{-4pt}
Reuse metrics include software-mention counts in publications~\cite{howison2016software,du2021software}, package-manager download counts, dependent-package counts, fork counts, and platform stars, easily automated from registries (e.g. PyPI, CRAN, conda-forge) and from publication indices that detect software mentions. They are particularly vulnerable to gaming and to confusion between \emph{visibility} and \emph{value}: a million downloads driven by transitive dependencies say little about whether a package advanced specific research.

\vspace{-4pt}
\subsection{Credit and Attribution (RQ1, RQ3, RQ4)}
\vspace{-4pt}
Credit metrics align with the FORCE11 Citation Principles~\cite{smith2016force11} and emerging standards including the Citation File Format~\cite{druskat2021cff}, DOIs minted through Zenodo--GitHub integration, and CRediT-style contributor role taxonomies~\cite{credit}. Indicators include software citation counts through Crossref, CITATION.cff presence and validity, and share of releases with persistent identifiers. Here the validation challenge is not metric definition but \emph{practice}: citation behavior remains inconsistent across disciplines~\cite{li2018software}, so absent citations reflect disciplinary norms rather than nonuse.

\vspace{-6pt}
\subsection{Community Health (RQ1, RQ4)}
\vspace{-4pt}
The CHAOSS project~\cite{chaoss} has produced guidance on community health metrics: contributor diversity, issue-response time, newcomer retention, governance maturity, and code-of-conduct adoption. These metrics surface social and organizational dimensions of research software that code-only analyses often miss. They also tend to undercount labor performed off-platform (e.g. mailing lists, working groups, private mentoring), and we return to that equity concern in \Cref{sec:implications}.

\vspace{-6pt}
\subsection{Security and Supply Chain (RQ1, RQ5)}
\vspace{-4pt}
Security-oriented metrics matter increasingly as research software pipelines incorporate many dependencies of unclear provenance. Indicators include known-vulnerability counts (CVE matching), dependency freshness, signed-release fraction, software bill of materials (SBOM) presence~\cite{cisa2023sbom}, and dependency review status through projects like OpenSSF Scorecard~\cite{scorecard}. These metrics automate well but can produce false positives in research contexts, where a dependency may be ``stale'' but stable.

\vspace{-6pt}
\subsection{Environmental and Performance (RQ1, RQ5)}
\vspace{-4pt}
A newer family of metrics tracks the energy and resource footprint of research software, including runtime, memory, FLOPs, and estimated energy consumption per workload~\cite{henderson2020greenml,patterson2021carbon}. These matter particularly for AI-era pipelines. They are difficult to compare across infrastructures~\cite{henderson2020greenml,patterson2021carbon}, and can incentivize relocating compute rather than reducing it. This family is the thinnest in the focused review relative to its growing importance: practice is \emph{established} for ML training workloads but \emph{emerging} for research software generally, and AI workloads are precisely what push energy accounting from a specialist concern toward a routine reporting need; standardized per-run energy reporting remains an open gap (\Cref{tab:ai-gaps}).

\vspace{-4pt}
\section{Evidence and Validation Gaps (RQ3, RQ4)}
\vspace{-4pt}
\label{sec:evidence}
A landscape of available metrics says little, on its own, about which metrics are valid for the constructs they claim to measure: a metric being widely \emph{used} is not the same as being \emph{valid}~\cite{hicks2015leiden,fenton2014software}. The focused review surfaced three observations, presented in turn below.
\vspace{-8pt}
\subsection{Indicators Are Often Confused with Constructs}
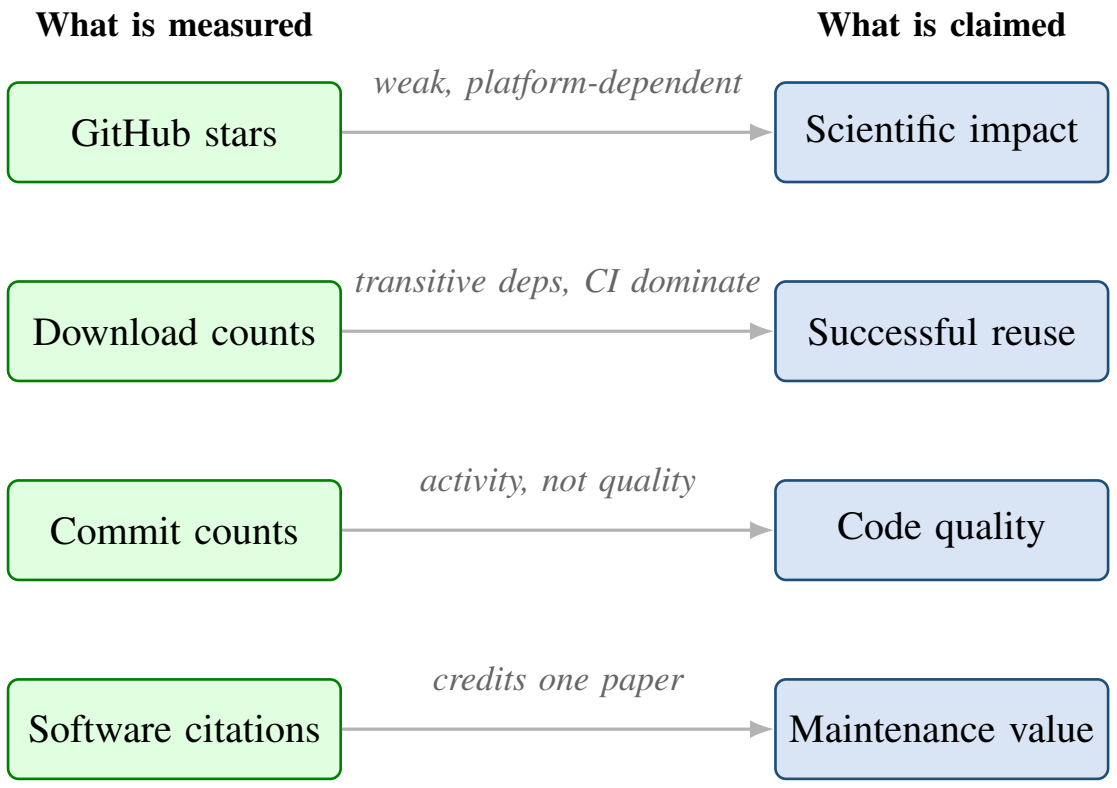
\begin{figure}[!t]
\centering
\begin{tikzpicture}[scale=0.94, transform shape,
    every node/.style={font=\footnotesize},
    indicator/.style={
        rectangle, draw=green!50!black, fill=green!12, rounded corners=2pt,
        line width=0.5pt, text width=22mm, minimum height=7mm,
        align=center, inner sep=2pt
    },
    construct/.style={
        rectangle, draw=takeawayBorder, fill=stageScreen!30, rounded corners=2pt,
        line width=0.5pt, text width=22mm, minimum height=7mm,
        align=center, inner sep=2pt
    },
    arrow/.style={-{Latex[length=1.6mm]}, semithick, draw=gray!60},
    gaplabel/.style={
        font=\scriptsize\itshape\color{black!60}, anchor=south
    },
    colhead/.style={font=\scriptsize\bfseries, anchor=south}
]

\def\indX{0}
\def\conX{5.4}
\def\midX{2.7}

\node[colhead] at (\indX,0.55) {What is measured};
\node[colhead] at (\conX,0.55) {What is claimed};

\node[indicator] (i1) at (\indX,0)      {GitHub stars};
\node[construct] (c1) at (\conX,0)      {Scientific impact};
\draw[arrow] (i1) -- (c1);
\node[gaplabel] at (\midX, 0.10) {weak, platform-dependent};

\node[indicator] (i2) at (\indX,-1.4)   {Download counts};
\node[construct] (c2) at (\conX,-1.4)   {Successful reuse};
\draw[arrow] (i2) -- (c2);
\node[gaplabel] at (\midX, -1.30) {transitive deps, CI dominate};

\node[indicator] (i3) at (\indX,-2.8)   {Commit counts};
\node[construct] (c3) at (\conX,-2.8)   {Code quality};
\draw[arrow] (i3) -- (c3);
\node[gaplabel] at (\midX, -2.70) {activity, not quality};

\node[indicator] (i4) at (\indX,-4.2)   {Software citations};
\node[construct] (c4) at (\conX,-4.2)   {Maintenance value};
\draw[arrow] (i4) -- (c4);
\node[gaplabel] at (\midX, -4.10) {credits one paper};

\end{tikzpicture}
\vspace{-6pt}
\caption{Common indicator--construct conflations in research software metrics.}
\label{fig:constructs}
\vspace{-5pt}
\end{figure}

\vspace{-4pt}
A recurring validity gap in the focused review is conflation of an indicator with the construct it is supposed to represent. We reserve frequency claims for the planned full mapping study (\Cref{sec:method}); the examples below illustrate the pattern rather than quantify it, and span several metric families:

\begin{itemize}
    \item GitHub stars are sometimes invoked as evidence of \emph{impact}, but stars more directly indicate \emph{visibility} and \emph{interest signaling}. The correlation with reuse or scientific outcomes is weak and platform-dependent~\cite{borges2018githubstars}.
    \item Download counts indicate \emph{access}, not successful \emph{reuse}; transitive dependencies, automated mirror traffic, and CI installs can dominate the totals.
    \item Commits indicate \emph{activity}, not \emph{quality}; AI-assisted commits plausibly compound this distortion \emph{(anticipated)}, although the magnitude of the effect awaits empirical study.
    \item Citation counts capture \emph{scholarly recognition} through papers that use or describe the software, not the \emph{maintenance value} of the software itself~\cite{howison2016software,li2018software}.
\end{itemize}
\noindent These conflations do not argue against the indicators, only against using them \emph{alone} to support strong construct claims. \Cref{fig:constructs} pairs each indicator with the construct it is often taken to represent.

\vspace{-4pt}
\subsection{Validation Evidence Is Uneven Across Families}
\vspace{-5pt}
The validation literature concentrates in a few families. Software-quality metrics from traditional software engineering have decades of empirical study \emph{(established)}, though systematic reviews find that the predictive power of structural measures such as cyclomatic complexity varies substantially across languages, contexts, and study designs~\cite{hall2012slr,radjenovic2013metrics}, and their applicability to research code in particular is contested~\cite{wilson2014best,wilson2017good}. FAIR metrics have an emerging validation literature concerned with scoring instruments and the relationship between scores and reuse~\cite{lamprecht2020fair} \emph{(emerging)}. By contrast, community-health metrics from CHAOSS~\cite{chaoss} are \emph{well defined} but their relationship to downstream outcomes (project survival, contributor retention, scientific reuse) is still being established \emph{(emerging)}. AI-era metrics for code provenance or generation share are largely \emph{anticipated}: proposed or adjacent to existing work but not yet empirically validated at scale, while reviewer-effort measurement now has established evidence at industrial scale, though not yet in research-software settings (\Cref{tab:ai-gaps}).
\vspace{-8pt}
\vspace{-7pt}
\subsection{Risks of Bias, Gaming, and False Precision}
\vspace{-4pt}
Several risks recurred in the reviewed sources. We list them as concrete patterns rather than abstract concerns:
\begin{itemize}
    \item \textbf{Platform bias.} Metrics computed solely from one platform advantage projects hosted there and systematically miss projects hosted elsewhere.
    Filtering decisions themselves materially shape project ranking~\cite{malviya2024filtering}.
    \item \textbf{Discipline bias.} Software citation practices vary substantially across disciplines, so cross-disciplinary comparison of citation-based metrics is hazardous~\cite{li2018software}.
    \item \textbf{Visibility bias.} Reuse and adoption metrics favor projects with mature documentation, citation infrastructure, and outreach, which themselves depend on institutional and RSE support~\cite{cosden2023rse,carver2022survey}.
    \item \textbf{Gaming.} Several metrics are manipulable (stars, downloads, commit cadence); AI assistance plausibly lowers the cost of gaming on at least the surfaces noted in \Cref{sec:discussion} (synthetic commits, generated tests, automated issue activity) \emph{(anticipated)}.
    \item \textbf{Invisible labor.} Maintenance, mentoring, support, code review, governance, and community care happen off the metric ledger entirely; contributor-role taxonomies such as CRediT and surveys of RSE practice make this gap visible but do not yet close it~\cite{credit,cosden2023rse,carver2022survey}.
    \item \textbf{False precision.} Decimal-point reporting of noisy indicators can lend unwarranted authority to weak signals~\cite{hicks2015leiden}.
\end{itemize}
\vspace{-5pt}
\begin{takeaway}
\emph{Anti-patterns to watch for.} A metric pipeline should be questioned if any of these apply: (1) single-source: one hosting platform, registry, or citation database; (2) single-number, with no panel, uncertainty, or qualitative context; (3) designed for a different purpose and repurposed without re-validation~\cite{fenton2014software}; (4) rewards the action you want without rewarding the outcome you want, the classic Goodhart-style failure mode~\cite{hicks2015leiden}; (5) those measured cannot audit or dispute it, in tension with
the 
principles of DORA and the Leiden Manifesto~\cite{dora,hicks2015leiden}.
\end{takeaway}

\vspace{-4pt}
\section{Discussion: Metrics in the Age of AI}
\vspace{-4pt}
\label{sec:discussion}
What does AI-assisted development change about how we should measure research software? As argued in \Cref{sec:introduction}, the need for evidence on reliability, reuse, maintainability, citation, security, and scientific trustworthiness does not disappear when production is partly automated. We organize this discussion around four pressures AI sharpens (provenance, maintainability, responsible impact, gaming), with \Cref{tab:ai-gaps} as a fuller inventory of the gap surface. 
Rows marked anticipated should be read as a research agenda derived from known metric failure modes and AI-assisted development trends, not as claims of observed widespread failure. We expand on each pressure in turn, distinguishing where evidence exists from where we offer informed projections, then turn to gaps that broaden what a metric system must cover, and close with three implications for the metric system.

\begin{table*}[tbp]
\caption{AI-era gaps in research software metrics and suggested directions.}
\vspace{-6pt}
\label{tab:ai-gaps}
\centering
\scriptsize
\setlength{\tabcolsep}{4pt}
\renewcommand{\arraystretch}{1.0}
\begin{tabularx}{\textwidth}{@{} p{0.28\textwidth} p{0.34\textwidth} p{0.34\textwidth} @{}}
\toprule
\textbf{Gap} & \textbf{Why current metrics fall short} & \textbf{Suggested direction} \\
\midrule
AI-generated code provenance~\cite{chen2021codex,smith2016force11,druskat2021cff} \emph{(anticipated)} &
Commits, blame, and authorship metadata assume human authorship; the tool or model that produced or rewrote code is not tracked &
Provenance metadata at commit and file granularity; model and prompt identifiers; signed attestations; automated detection \\
Maintainability of AI-assisted code~\cite{vaithilingam2022expectation,perry2023copilot,kabir2024llm} \emph{(emerging)} &
Static-analysis metrics may pass on AI code that is nonetheless brittle, opaque, or hard to extend &
Maintainability metrics that include change-failure rate, time-to-modify, and reviewer effort, not only structural complexity \\
Human review effort~\cite{perry2023copilot,vaithilingam2022expectation,radar26,RevRisk26,aiCodeReview26} \emph{(established at industrial scale)} &
Code-review counts treat all reviews as equivalent; AI-generated code shifts effort onto reviewers, and industrial deployments now offload low-risk review to automated triage while reserving human attention for high-risk diffs~\cite{radar26,RevRisk26} &
Reviewer-effort indicators (time, depth, follow-up changes, auto- vs.\ human-approved share) calibrated for AI-assisted contributions \\
AI-generated test inflation~\cite{kabir2024llm,vaithilingam2022expectation,fenton2014software} \emph{(anticipated)} &
Test coverage and CI pass rates assume tests reflect human reasoning about expected behavior; AI-generated tests may exercise coverage without checking the property the developer cared about &
Behavioral assertions and mutation-testing scores alongside coverage; review of test intent, not just test count \\
Maintainer-activity inflation~\cite{thakur2025abandoned,cosentino2015busfactor,vaithilingam2022expectation} \emph{(anticipated)} &
Maintainer-activity metrics (commit cadence, PR throughput) assume commits reflect human effort; AI-assisted commits inflate the signal while review effort grows &
Pair raw commit activity with review-effort signals (time-to-merge, reviewer comments per PR) for a fuller view of sustainability \\
Model / data dependency chains~\cite{cisa2023sbom,scorecard,patterson2021carbon} \emph{(emerging)} &
Software dependency graphs ignore model and training-data dependencies that materially affect reproducibility &
Extend SBOMs to ML/AI: model bills of materials, dataset bills of materials \\
Reproducibility of AI workflows~\cite{larsonneur2018workflows,grayson2023workflows,lamprecht2020fair} \emph{(emerging)} &
Container and environment metrics omit non-deterministic AI components and external API dependencies &
Reproducibility metrics that include determinism level, model-version pinning, and external-call recording \\
Energy and resource cost~\cite{henderson2020greenml,patterson2021carbon} \emph{(established for ML; emerging for general software)} &
Performance metrics rarely capture energy, and AI workloads materially shift the cost profile of compute &
Standardized energy and compute reporting per pipeline run; carbon-aware indicators \\
Reuse-metric inflation~\cite{howison2016software,borges2018githubstars,du2021software} \emph{(anticipated)} &
Download counts, dependent-package counts, and mirror traffic assume human consumption; CI pipelines, automated mirrors, and transitive AI-driven dependency installs increasingly dominate the totals &
Distinguish human from automated downloads where possible; report dependent-project counts and named adoption alongside raw download totals \\
Community-health signal pollution~\cite{chaoss,cosden2023rse,kabir2024llm} \emph{(anticipated)} &
Issue-response time and discussion activity assume human responses; bot-driven and AI-assisted replies inflate responsiveness without reflecting community capacity &
Track human-reviewed responses separately from automated ones; report reviewer-effort indicators alongside response-time medians \\
Attribution of RSE labor~\cite{cosden2023rse,credit,druskat2021cff,carver2022survey} \emph{(problem established; remedies emerging)} &
Authorship and commit metrics overweight code-producing activity; non-code RSE work in review, design, testing, and mentoring goes uncounted &
Contributor-role metrics (CRediT-style) that recognize non-code labor; review and mentoring as first-class signals \\
Human versus generated contribution~\cite{chen2021codex,kabir2024llm,ieeeai} \emph{(emerging / policy-driven)} &
No standard way separates human authorship from generated text in metrics &
Authorship attestations and policy disclosures linked to contributions \\
\bottomrule
\end{tabularx}
\end{table*}
\vspace{-4pt}

\subsection{AI Raises the Bar for Provenance Metrics (RQ5)}
\vspace{-4pt}
AI-assisted contributions disturb the standard assumption that a commit author wrote, or even reviewed in detail, the code they have pushed. As AI-assisted tooling spreads in research-software workflows, the relevant provenance questions extend beyond \emph{who} to include \emph{which model}, \emph{which prompt}, \emph{which tools}, and \emph{which level of human review}~\cite{chen2021codex,perry2023copilot,vaithilingam2022expectation}. Existing software citation and FAIR metadata frameworks~\cite{smith2016force11,druskat2021cff,barker2022fair4rs} could extend to record this provenance, but the extensions are not yet standard. Conference and publisher policies mandating author-level AI-use disclosure~\cite{usrse2026call,ieeeai} are a useful precedent, but they sit at the manuscript layer and do not yet translate into commit- or release-level metadata.

\vspace{-4pt}
\subsection{AI Raises the Bar for Maintainability Metrics (RQ5)}
\vspace{-4pt}
Early AI-coding evaluation studies report that code which passes initial tests may still be brittle, opaque, or hard to modify by future maintainers~\cite{vaithilingam2022expectation,perry2023copilot}, properties that structural metrics such as cyclomatic complexity and churn can miss. This is not only an AI-era concern: the software measurement literature has long debated whether structural metrics reliably predict maintenance effort or defect rates, with predictive power varying substantially across languages, contexts, and study designs~\cite{hall2012slr,radjenovic2013metrics}. The indicator--construct gap identified in \Cref{sec:evidence} applies here too: cyclomatic complexity is an indicator of structural complexity, not a validated measure of actual maintainability. AI-generated code sharpens this existing gap because output that scores well on structural indicators can still be hard for human maintainers to understand or extend. Maintainability metrics derived from change-failure rates, time-to-modify after a regression, and reviewer-effort signals may better capture the long-tail maintenance cost resulting from AI-accelerated production \emph{(anticipated)}.

\vspace{-6pt}
\subsection{AI Raises the Bar for Responsible Impact Metrics (RQ5)}
\vspace{-4pt}
More code does not automatically mean more scientific value. Code volume, adoption, reuse, trust, contribution, maintenance, and scientific impact are distinct constructs, and \Cref{sec:evidence} illustrated how the literature can conflate them. In an AI era where producing code requires less effort, we should resist treating code production as a proxy for impact. This is the central argument for the responsible-metrics stance adopted from DORA~\cite{dora} and the Leiden Manifesto~\cite{hicks2015leiden}: as production becomes easier, the case for mixed-method, context-sensitive evaluation grows stronger.

\vspace{-4pt}
\subsection{AI Raises the Risk of Gaming Metrics (RQ5)}
\vspace{-4pt}
Several research-software metrics are trivially gameable today, and the marginal cost of gaming plausibly drops with AI assistance. Synthetic commits, generated documentation that looks complete but is not useful, superficial tests that exercise coverage without checking behavior, and automated issue activity all become cheaper at scale, so metric systems that worked when production was costly will need re-validation where it no longer is. The risk spans families: quality (superficial tests), community health (automated replies), and credit (AI-generated co-author lists). We present these as plausible gaming surfaces motivating periodic re-validation, not as observed and quantified failure modes, which would require the measurement program \Cref{sec:method} reserves for follow-on work.

\vspace{-4pt}
\subsection{AI Broadens What Must Be Measured (RQ5)}
\vspace{-4pt}
The four pressures above sharpen existing metric families; other rows of \Cref{tab:ai-gaps} broaden what a metric system must cover at all. \emph{Model and data dependency chains} \emph{(emerging)}: SBOMs~\cite{cisa2023sbom,scorecard} capture code dependencies but ignore the model weights, training corpora, and external inference APIs on which AI-era pipelines materially depend; extending the bill-of-materials idea to models and datasets is the natural direction. \emph{Reproducibility of AI workflows} \emph{(emerging)}: container metrics~\cite{larsonneur2018workflows,grayson2023workflows} certify that an environment rebuilds, but not whether non-deterministic model components or version drift in external APIs undermine the result, so reproducibility indicators need determinism level, model-version pinning, and external-call recording. \emph{Human review effort} \emph{(established at industrial scale)}: review counts treat all reviews as equivalent, yet AI-generated contributions transfer effort from writing to reviewing rather than reducing it overall, and industrial-scale deployments now route low-risk diffs through automated triage so scarce human review can concentrate on high-risk changes~\cite{perry2023copilot,vaithilingam2022expectation,radar26,RevRisk26,aiCodeReview26}; effort-weighted review signals and the auto- vs.\ human-approved share would make a growing, currently invisible category of RSE labor legible in evaluation.

A further broadening is specific to \emph{research} software. As code generation becomes cheap, the differentiating question is increasingly whether software produces trustworthy scientific results. Constructs such as verification, validation, and uncertainty quantification (V\&V/UQ)~\cite{oberkampf2010vvuq} are central to that trust, yet under-served by the eight families, largely because the literature feeding this review is dominated by general software engineering sources. Mapping V\&V/UQ practice to constructs, indicators, and metrics is a priority direction for the research-software-specific extension of this landscape (\Cref{sec:conclusion}).

\vspace{-4pt}
\subsection{Implications for the Metric System}
\vspace{-4pt}
Three implications follow. First, metric pipelines need provenance data sources: commit metadata, model identifiers, and human-review attestations as first-class inputs, extending the software-citation and FAIR metadata frameworks that already exist~\cite{smith2016force11,druskat2021cff,barker2022fair4rs}. Second, metric systems should report \emph{ensembles}, not single numbers, and expose enough detail for verification by those being evaluated, the transparency stance shared by DORA and the Leiden Manifesto~\cite{dora,hicks2015leiden}. Third, evaluation frameworks should be reviewed periodically for new gaming surfaces as new AI tools enter the workflow. 
We translate these implications into stakeholder-specific guidance next.
\vspace{-4pt}

\vspace{2pt}
\section{Implications for RSEs and 
Collaborators}
\vspace{-4pt}
\label{sec:implications}
We translate the synthesis into practical guidance organized by stakeholder, condense it into a compact use-case decision guide (\Cref{tab:decision-guide}), and close with five questions practitioners can apply before adopting any candidate metric.

\begin{table}[!t]
\caption{Decision guide: common use cases mapped to the metric families, and the single-metric shortcut to avoid.}
\vspace{-6pt}
\label{tab:decision-guide}
\centering
\scriptsize
\renewcommand{\arraystretch}{1.1}
\begin{tabularx}{\columnwidth}{@{}>{\raggedright\arraybackslash}p{0.27\columnwidth} X >{\raggedright\arraybackslash}p{0.2\columnwidth}@{}}
\toprule
\textbf{Use case} & \textbf{Draw from families} & \textbf{Avoid alone} \\
\midrule
Continue, expand, or retire a package & Quality; community health; reuse; qualitative user evidence & Stars; download counts \\
RSE-group portfolio review & Sustainability; credit (incl.\ non-code labor); community health & Commits or LOC per RSE \\
Funder proposal or renewal & Sustainability plan; FAIR / citation infrastructure; security & Star or download promises \\
Reviewing an AI-assisted contribution & Quality (test intent); credit / provenance; reviewer effort & CI pass + commit volume \\
Promotion or hiring case & Credit; reuse; community health; qualitative impact cases & Publication citations \\
\bottomrule
\end{tabularx}
\end{table}

\vspace{-4pt}
\subsection{Implications by Stakeholder}
\vspace{-4pt}
\paragraph{For the Developer Community (RSEs)}
First, start from the question, not the dashboard: define the construct of interest (``Is this library reusable by external groups?'') before choosing metrics, so selection from \Cref{tab:families} is question-driven, not catalog-driven. Second, pair structural with social signals: combine code-level indicators (tests, CI, FAIR compliance) with community-level indicators (contributor diversity, issue response, governance) to avoid over-weighting the easily automated signals flagged in \Cref{sec:evidence,sec:discussion}. Third, surface invisible labor: use contributor-role taxonomies (CRediT, all-contributors, CITATION.cff)~\cite{credit,druskat2021cff} to record review, mentoring, documentation, and user support alongside code authorship. Finally, record AI involvement where tooling allows (commit trailers, signed attestations, release-note disclosures); broader commit- and file-level provenance remains an \emph{anticipated} capability (\Cref{sec:discussion}).
\paragraph{For Principal Investigators and Team Leads}
Choose a small, justified metric panel: three to five metrics from different families, with stated rationale, are easier to defend than long lists of platform counts and align with the responsible-metrics stance of DORA and the Leiden Manifesto~\cite{dora,hicks2015leiden}. Interpret in context: a bus factor of one is not always bad, and a scope-stable small project may not need ten maintainers; state the context, then the number. Avoid single-platform dependence: if all metrics come from one hosting platform, you are measuring that platform as much as the software (\Cref{sec:evidence}~\cite{malviya2024filtering}).
\paragraph{For Institutions}
Adopt a responsible-metrics policy. DORA and the Leiden principles~\cite{dora,hicks2015leiden} were drafted for publication-based assessment, but their stances on responsible use, transparency, and contextual interpretation transfer to research-software evaluation in promotion, hiring, and awards, with software-specific extensions from FAIR4RS~\cite{barker2022fair4rs} and the FORCE11 principles~\cite{smith2016force11}. Invest in attribution infrastructure: persistent identifiers, CITATION.cff files, and Zenodo deposits change what is countable, and institutions can lower the cost of generating them. Recognize maintenance and stewardship: sustained maintenance of community-relied software should be a legible career activity, not an invisible one~\cite{mundt2023public}.

\paragraph{For Funders}
Ask for metric plans, not metric promises: require applicants to specify which metrics they will use, why, and how they will interpret weak or contested signals. Fund the data infrastructure, not only the tooling; persistent identifiers, software-citation indexing, and software-mention corpora are public goods that enable everyone else's metrics. Resist single-score evaluation: no single number reliably summarizes research software value.

\vspace{-6pt}
\subsection{Five Questions Before Adopting a Metric}
\vspace{-4pt}
Drawing the above together, we propose five questions to apply to any candidate metric before adoption; they complement the \Cref{sec:evidence} anti-patterns, which flag when an existing pipeline should be questioned, by structuring the choice of a new one.
\begin{enumerate}
    \item \emph{Purpose.} What decision will this metric inform, and for which stakeholder?
    \item \emph{Construct.} What is the underlying property of interest, and is this metric a credible indicator of it?
    \item \emph{Validity.} What evidence supports the link between the metric and the construct in contexts like ours?
    \item \emph{Risks.} What biases, gaming surfaces, or equity concerns apply, and how will we monitor for them?
    \item \emph{Composition.} What other metrics, qualitative context, or human judgment will accompany this one?
\end{enumerate}

\smallskip
\noindent\emph{Five Questions in action.} An RSE wants to make the case to a PI that their simulation library deserves continued investment. \textit{Purpose}: argue for continued support based on real usage, not lines of code. \textit{Construct}: scientific reuse beyond the host lab. \textit{Validity evidence}: combine software-citation counts~\cite{smith2016force11} with downstream paper mentions and an external-reuse count from publication metadata. \textit{Risks}: visibility bias and citation-discipline norms if the library is interdisciplinary. \textit{Composition}: pair the quantitative panel with two qualitative items: three named external groups using the library, and a short case where the library mattered. The result is a small panel with stated assumptions, not a single score.

\smallskip
\noindent\emph{Five Questions in action: AI-assisted code review.} Consider an RSE evaluating a maintenance branch where the contributor used an AI pair-programmer. A conventional panel of commit volume, passing CI, and added coverage would credit the branch as healthy. The five questions shift the evaluation: \textit{Purpose}: assess design judgment, not commit volume. \textit{Construct}: which commits represent human design decisions versus accepted AI suggestions. \textit{Validity}: passing CI and test coverage carry weaker construct validity once tests can be AI-generated~\cite{kabir2024llm}. \textit{Risks}: review-effort burden on senior maintainers may grow and stay invisible, an \emph{anticipated} pattern consistent with early AI-coding evaluation literature~\cite{vaithilingam2022expectation,perry2023copilot}. \textit{Composition}: the evaluation requires AI-disclosure metadata, reviewer-effort signals, and qualitative assessment of the design changes, none currently captured by conventional panels.
\vspace{-4pt}
\begin{takeaway}
\emph{
\noindent We do not advocate a universal ``Research Software Impact Score.'' We argue, with DORA~\cite{dora} and the Leiden Manifesto~\cite{hicks2015leiden}, that metrics should be selected according to purpose, interpreted with qualitative context, made transparent, and reviewed for bias and unintended incentives.}
\end{takeaway}
\vspace{-4pt}

\vspace{-4pt}
\section{Threats to Validity}
\label{sec:threats}
\subsection{Generalizability}
\vspace{-4pt}
The corpus is in English, biasing toward Anglophone communities, and the search channels are mostly open-science venues, so the corpus weights fields with active RSE engagement and underrepresents research software outside those venues. Our claims say \emph{what the focused review identifies}, not \emph{what every project should do}.
\vspace{-4pt}
\subsection{Construct Validity}
\vspace{-4pt}
The family structure reflects coding decisions that necessarily simplify a heterogeneous literature; some families (e.g. sustainability and community health) partially overlap but were kept distinct because they draw on different data sources and serve different stakeholders. 
Researcher bias from seeding the focused review with the author team's working knowledge is a related threat.
The family structure remains an interpretive contribution.

\vspace{-4pt}
\subsection{Internal Validity}
\vspace{-4pt}
The principal internal-validity caveat is scope. The landscape, evidence assessment, and AI-era gap analysis in this paper rest on a focused review of 49 substantive references rather than a completed full mapping pass. The 727-record triaged corpus accompanying the paper is an artifact prepared for that future pass, not the evidentiary basis for the families described in \Cref{sec:taxonomy}.
Current SE mapping-study guidelines~\cite{kitchenham2007guidelines,petersen2015guidelines,wohlin2012experimentation} recommend two-reviewer screening with Cohen's $\kappa$; the present focused review sits upstream of that pass, which the planned mapping study adopts. A secondary threat is that several included sources are community reports rather than peer-reviewed research; we treat these as community vocabulary, and the \emph{established / emerging / anticipated} evidence-tier labels in \Cref{sec:evidence,sec:discussion} signal where community sources carry the validation argument and where peer-reviewed evidence does.

\vspace{-4pt}
\subsection{Conclusion Validity}
\vspace{-4pt}
This paper presents a landscape synthesis,
not a meta-analysis, so we
draw no quantitative conclusions about the prevalence or effectiveness of any individual metric. Our claims are structural, 
and the
\emph{established / emerging / anticipated} evidence-tier labels that run through \Cref{sec:evidence,sec:discussion} mark the boundary between synthesis of the focused review and author-team interpretation.

\vspace{-2pt}
\section{Conclusion}
\label{sec:conclusion}
\vspace{-4pt}
We have argued that research software metrics matter increasingly in the age of AI, and that the literature defining them remains fragmented. This paper organized the research software metrics landscape from a focused review of 49 references and accompanies the synthesis with a reproducibility artifact.
The focused review identifies eight recurring metric families (RQ1) drawing on shared data sources that create both consolidation opportunity and platform-dependence risk (RQ2); finds that a recurring validity gap is conflation of indicators with their constructs (RQ3); observes validation evidence unevenly distributed across families (RQ4); and identifies AI as sharpening provenance, maintainability, impact, and gaming pressures while widening gaps around
dependencies, 
reproducibility, signal inflation, and credit for human labor that AI does not perform (RQ5). The core practitioner guidance is: 
choose metrics by purpose, combine them, interpret in context, and review for bias.
\textit{Future work} proceeds on three fronts. First, a full systematic mapping study against the 727-record corpus with two-reviewer screening and Cohen's $\kappa$, adding the prevalence, coverage, and agreement evidence the focused review reserves. Second, a practitioner survey of RSEs, PIs, institutions, and funders to validate the eight-family landscape against practice, probe research-software-specific constructs such as verification, validation, and uncertainty quantification (\Cref{sec:discussion}), and pilot the Five Questions checklist with RSE groups; the survey also implements the member-check mechanism of \Cref{sec:threats}. A companion thread will couple the eight families to a program-evaluation logic model~\cite{kellogg2004logic}, anchoring metric selection to the stage of the research-software value chain (inputs, activities, outputs, outcomes, impacts) a stakeholder cares about. Third, metrics that AI \emph{enables} rather than only threatens: software-mention mining and semantic reuse detection could address the known inadequacy of citation-based impact signals~\cite{howison2016software}; candidate journal-venue threads include empirical AI-era evidence from a prior catalog of scientific software repositories~\cite{thakur2023scicat}. The contribution throughout is not a universal score but a practical guide for choosing and interpreting research software metrics responsibly in AI-era science.

\balance

\vspace{-2pt}
\section*{Acknowledgments}
\small
\noindent We thank the US-RSE community, the FAIR4RS community, the CHAOSS project, and the RSQKit collaboration whose work informed this synthesis.

We acknowledge the support of the Consortium for Open-Source Research Software Advancement (CORSA), a project supported by the U.S. Department of Energy, Office of Science, Office of Advanced Scientific Computing Research, Next-Generation Scientific Software Technologies program, under contract number DE-AC05-00OR22725.

This manuscript has been authored by UT-Battelle, LLC, USA under Contract No.~DE-AC05-00OR22725 with the U.S. Department of Energy. The publisher, by accepting the article for publication, acknowledges that the U.S. Government retains a nonexclusive, paid up, irrevocable, worldwide license to publish or reproduce the published form of the manuscript, or allow others to do so, for U.S. Government purposes. The DOE will provide public access to these results in accordance with the DOE Public Access Plan (\url{https://www.energy.gov/downloads/doe-public-access-plan}).

Per USRSE'26~\cite{usrse2026call} and IEEE~\cite{ieeeai} AI-assisted-writing policies, the authors used OpenAI ChatGPT, Anthropic Claude, and Google Gemini for language/LaTeX editing and search-triage code assistance; the authors reviewed and edited all AI-assisted content and take full responsibility.

\clearpage

\bibliographystyle{myIEEEtran}
\bibliography{references}

\clearpage
% \appendix

% \section*{Statement of Connection to the US-RSE Community}
% \label{sec:usrse-connection}
% \input{connection-statement}

\end{document}